\documentclass[10pt,prb,aps,superscriptaddress,amsmath,amssymb,twocolumn,floatfix,longbibliography]{revtex4-1}
\usepackage{mathtools}
\usepackage{mathrsfs, amsmath, wasysym}
\usepackage{bbold}
\usepackage{dsfont}
\usepackage{braket,mleftright}
\usepackage[mathscr]{eucal}
\usepackage{xfrac}
\usepackage{graphicx}
\usepackage{dcolumn}
\usepackage{bm}
\usepackage{manfnt}
\usepackage{color}
\usepackage{tabularx}
\usepackage{natbib}
\usepackage[bookmarksopen, colorlinks,linkcolor=blue,citecolor=blue,urlcolor=blue]{hyperref}
\usepackage[normalem]{ulem}
\usepackage[all]{hypcap}
\usepackage[dvipsnames,usenames,table]{xcolor}

\newcommand{\bg}{\begin{gather}}
\newcommand{\eg}{\end{gather}}

\newcommand{\be}{\begin{equation} }
\newcommand{\ee}{\end{equation}}
\newcommand{\beq}{\begin{eqnarray}}
\newcommand{\eeq}{\end{eqnarray}}

\newcommand{\br}{{\bf r}}
\newcommand{\bR}{{\bf R}}

\begin{document}

\title{Parametric neural-network wavefunctions across a bosonic phase transition
}

\author{Daniil S. Antonenko}
\email{daniil.antonenko@yale.edu}
\affiliation{Department of Physics, Yale University, New Haven, Connecticut 06520, USA}
\author{Liang Fu}
\affiliation{Department of Physics, MIT, Cambridge, USA}
\author{Leonid I. Glazman}
\affiliation{Department of Physics, Yale University, New Haven, Connecticut 06520, USA}
\date{\today}

\begin{abstract}

We propose \textit{Boseformer}, a universal attention-based neural-network architecture for representing the wavefunctions of particles with bosonic quantum statistics, including superfluid and crystal phases.
The model is trained to generate the wavefunction as a function of a Hamiltonian parameter across the quantum phase transition between these phases.
Once trained with variational Monte Carlo optimization, it can serve as a foundation model describing the phase transition without the need to retrain it for individual parameter values.
We demonstrate our method on a system of two-dimensional bosons with $1/r^3$ repulsive interaction.

\noindent\textbf{Code availability:} \url{https://github.com/andanya/boseformer}

\end{abstract}

\maketitle
\section{Introduction}

Recently, neural variational wavefunctions, or neural quantum states (NQS), have emerged as powerful ansätze capable of approximating a wide variety of quantum states~\cite{Carleo2017-NQS}. Typically paired with a variational Monte Carlo (VMC) optimization protocol, they have been successfully applied to a number of electronic and spin-lattice models and to continuum systems in quantum chemistry~\cite{Pfau2020-FermiNet,Psiformer} and condensed matter physics~\cite{Cassella2023-DiscoveringPhaseTransitions,Teng2025-SelfAttentionFQH, Geier2025-SelfAttention}. 

The study of bosonic systems with NQS has received less attention (see, however, Refs.~\onlinecite{Pescia2022-PeriodicBosons,Freitas2023-HeliumClusters,Linteau2025-HeliumPhaseDiagram}). The antisymmetry of fermionic wavefunctions is often ensured by using determinants of generalized orbitals~\cite{Pfau2020-FermiNet,Psiformer}.  
A naive generalization of this method to bosonic systems would use a permanent in place of a determinant, which is numerically infeasible: computing a permanent is much harder, requiring $O(N2^N)$ operations, compared with $O(N^3)$ for a determinant~\cite{Ryser1963}. An alternative approach is to use a permutation-invariant network that produces symmetric outputs, as implemented in the DeepSets architecture~\cite{Zaheer2017-DeepSets,Pescia2022-PeriodicBosons}.

\begin{figure}
    \centering
    \includegraphics[width=0.98\linewidth]{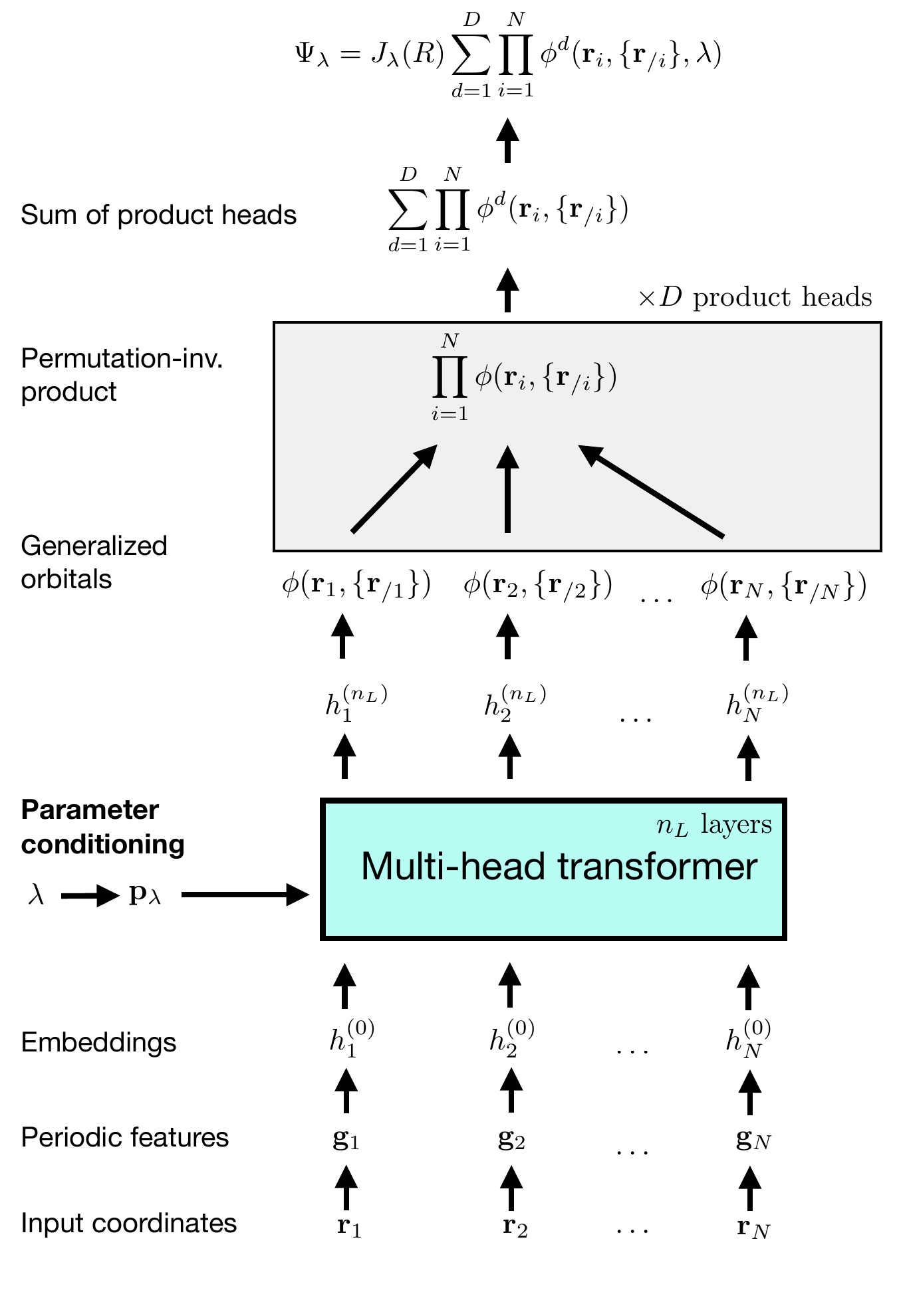}
    \caption{Architecture of the neural network representing the many-body wavefunction of $N$ bosonic particles. The coordinates of each particle, $\mathbf{r}_i$, and some Hamiltonian parameters, $\lambda$, are the model inputs. The streams corresponding to each particle exchange information in an equivariant way, see Sec.~\ref{sec:architecture}. As a result, the network outputs a permutation-invariant wavefunction representing bosonic statistics.
    \label{fig:architecture}
    }
\end{figure}

In this work, we introduce \textit{Boseformer}, a different, Transformer-based architecture~\cite{Vaswani2017-Attention} for representing the full many-body wavefunction of bosons, and demonstrate its advantages.
We use the product of generalized orbitals to generate the many-body bosonic wavefunction dependent on all particle coordinates; see Fig.~\ref{fig:architecture}. 
As we show below, our architecture is naturally capable of representing a superfluid state and also successful in describing crystalline phases of bosons. 

Our next major advancement is developing a parametric (foundation) model architecture~\cite{Zhang2023-TransformerQuantumState,Zaklama2026-LargeElectronModel,Nazaryan2026-QERNEL} that uses a single network to produce wavefunctions conditioned on the Hamiltonian parameters, which are treated as inputs to the network. In this setup, the network is optimized to find the ground states simultaneously on a selected set of parameter points. After training, the network predicts many-body wavefunctions over a continuous range of  Hamiltonian parameters without the need to re-train. Remarkably, we show that this approach works even for wavefunctions across a quantum phase transition. 

\begin{figure*}
    \centering
    \includegraphics[width=0.95\linewidth]{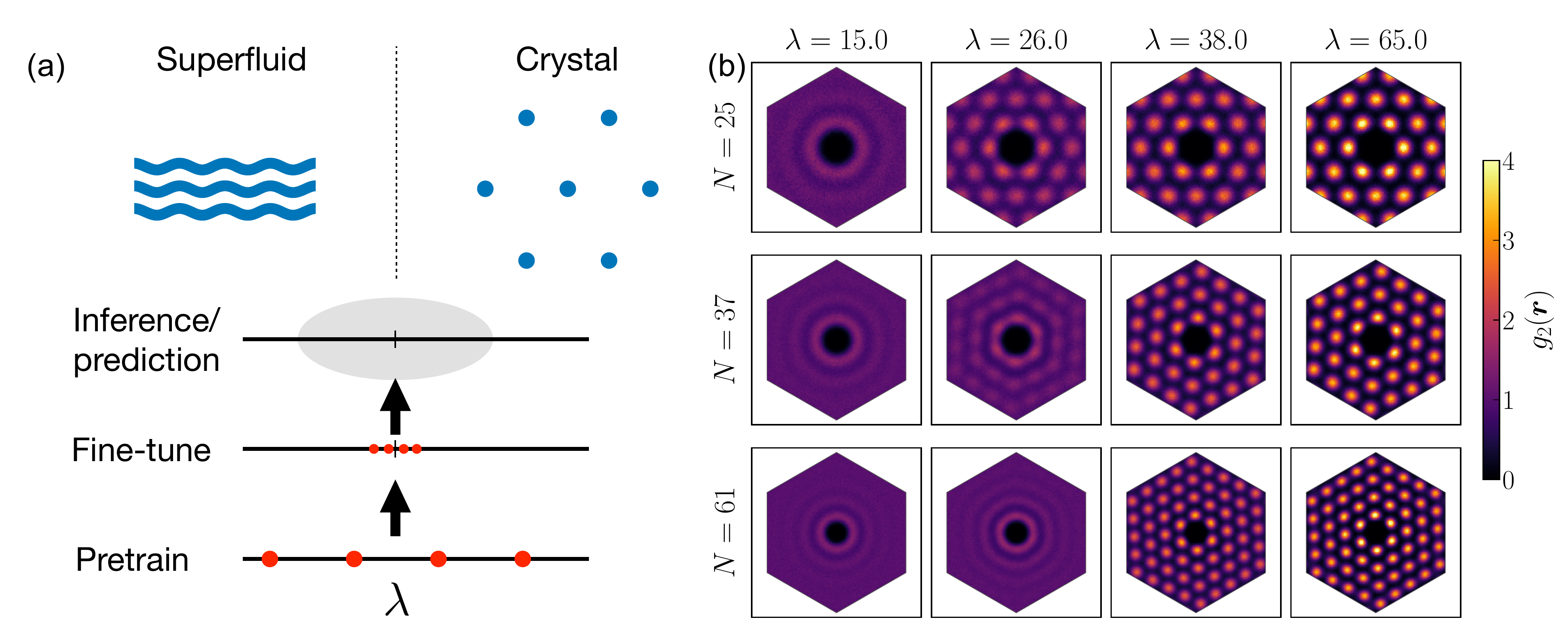}
    \caption{(a) Workflow of  \textit{Boseformer}: the network is first trained by minimizing the variational energy at a set of chosen interaction parameters $\lambda$ in the region of interest; then, it is fine-tuned to the vicinity of a phase transition or another interesting domain; finally, it can be used to predict wavefunctions in a region of parameters without the need to do variational optimization at each $\lambda$ individually. 
    (b) Pair correlation function $g_2(\bm{r})$ across the superfluid-to-crystal phase transition of $N$ two-dimensional bosons with $1/r^3$ repulsion. Every row was generated by running inference with a single pretrained model. 
    }
    \label{fig:g2-maps}
\end{figure*}

We demonstrate our method by applying it to the system of two-dimensional bosons with dipole-dipole repulsive interaction between them. Our model successfully predicts wavefunctions across the superfluid-to-crystal transition in this system; see Fig.~\ref{fig:g2-maps}. The energies of the obtained variational wavefunctions match those of the standard Diffusion Monte Carlo (DMC) method, typically used to study such systems. At the same time, our method, unlike DMC, produces the wavefunctions explicitly, which allows one to compute any ground-state observable.

\section{Physical system}

We study a continuum model of $N$ bosonic particles in two dimensions with a parabolic dispersion and pairwise repulsion. 
In general, as the interaction between the bosons is increased, the superfluid condensate is depleted and the importance of quantum correlations increases, as was studied in connection with helium in Refs.~\onlinecite{PenroseOnsager1956,Giorgini1996}. For the long-ranged interaction, a crystalline phase can become energetically competitive with the superfluid~\cite{Astrakharchik,Buchler2007,Mora2007}. 

For our study, we focus on the $1/r^3$ repulsion, relevant, for example, for the dipole--dipole interactions between out-of-plane excitons in TMD bilayers~\cite{BermanKezerashvili2016,Li2020DipolarExcitons}. The Hamiltonian in the units $\hbar = m =1$ reads:
\be \label{Hamiltonian}
    \mathcal{H} = -\frac{1}{2}\sum_{i=1}^N \bm{\nabla}_i^2 + \frac{\lambda}{\sqrt{\pi n}}  \sum_{i<j}  \frac{1}{|\br_i - \br_j|^3}.
\ee
We chose to write the interaction strength here as $\lambda/\sqrt{\pi n}$ for the following reason. A simple dimensional analysis shows that the kinetic energy per particle scales with the particle density as $n$, while the interaction energy scales as $n^{3/2}$ at fixed physical dipolar interaction strength. Upon the increase of density, the interaction energy grows and becomes comparable with the kinetic energy at some, {\sl independent of} $n$, value of $\lambda$. Previous study of dipolar bosons with the DMC method\cite{Astrakharchik, Mora2007} revealed a first-order phase transition from the superfluid to triangular crystalline phase at $\lambda \approx 30$.  Note that interaction dominates and leads to the formation of a crystalline phase at high densities. That distinguishes our system from one with Coulomb interaction, where the Wigner crystal formation occurs in the opposite limit of low density.

In the following, we will consider torus geometry with periodic boundary conditions. To restore translational invariance, the interaction term should be promoted to the sum over all periodic images to ensure translational invariance:
\be
    U = \frac{\lambda}{2\sqrt{\pi n}} \sum_{\bR} \sum_{i \neq j} \frac{1}{|\br_i - \br_j + \bR|^3 },
\ee
where $\bR$ is summed over the lattice defined by the simulation unit cell. Technically, this sum is performed via Ewald summation \cite{Ewald1921, NijboerDeWette1957, Mazars2011}. 

To minimize the finite-size effects, we make the simulation unit cell compatible with the geometry of the triangular crystal phase that forms at strong coupling. Specifically, we study a rhombus defined by the primitive lattice vectors $\bm{a}_1 = L(1,0)$ and $\bm{a}_2 = L(1/2, \sqrt{3}/2)$. 
We find it convenient to scale its size with the number of modeled  particles $N$ so that their density remains constant at the value of $n = 1/\pi$. That ensures consistent thermodynamic limits in the superfluid and crystal phases. 

Since there is no periodic potential in the model Hamiltonian \eqref{Hamiltonian}, the commensurability between the simulation and the crystal unit cells is determined solely by the arithmetic commensurability criterion. Specifically, rhombic unit cell is commensurate with the triangular lattice when the number of particles is one of the L{\"o}schian numbers \cite{Conway1999, Connelly2014},
\be \label{Loschian-numbers}
    N = m^2 + m n + n^2, \qquad m, n \in \mathds{Z},
\ee
which we will use in the computations. 

Interparticle interaction in \eqref{Hamiltonian} becomes singular when $\br_i$ becomes close to $\br_j$ for some $i, j$, which can create hurdles for numerical methods. The standard solution is to include Jastrow  factors~\cite{Jastrow1955} into the variational wavefunction, which allows to cancel the most singular part of the interaction and avoid spending neural network capacity on carving the correlation hole at close particle positions. To obtain a suitable Jastrow prefactor, we use the solution of the zero-energy two-body scattering problem for particles interacting with a $\lambda/r^3$ potential in two dimensions, $f_2(r) = K_0 (2\sqrt{\lambda / r})$~\cite{Astrakharchik}. Factoring out $f_2(r)$ from a two-body wavefunction generates Jastrow's kinetic contribution, $-\nabla^2 \log f_2 - (\nabla \log f_2)^2$, which cancels the leading $\lambda/r^3$ divergence as $r \to 0$. The exact combination that we use as a Jastrow factor reads
\be \label{Jastrow}
    J_{\lambda} (\bm{r}_1, \dots, \bm{r}_N) = \prod_{i<j} K_0\left(2\sqrt{\frac{\lambda}{\rho_{ij}}} \right),
\ee
where $\rho_{ij}$ is the smooth periodic distance introduced in Sec.~\ref{sec:architecture}. At small interparticle distances, it is close to $|\bm{r}_i - \bm{r}_j|$, while being smooth in the entire periodic unit cell ensuring seamless processing of the Laplacian in numerics.

\section{Architecture of the neural network}
\label{sec:architecture}

In this Section, we describe the details of the neural-network architecture that was used in our work. It takes as input a set of particle coordinates, $R = (\bm{r}_1, \dots,  \bm{r}_N)$, and the interaction strength $\lambda$. As the output, it produces the logarithm of the wavefunction amplitude, $\log \psi$. Due to the bosonic statistics, $\psi_\lambda(R)$ should be invariant under an arbitrary permutation of particle coordinates, which is achieved by using an equivariant neural network.

In the following, we present for concreteness  representations for two-dimensional particles as we focus on 2D in this work; generalization to other dimensions is straightforward.

We start with generating periodic input features for each particle (indexed with $i$) to account for the torus geometry:
\begin{subequations}
\begin{align}
    g_{i}^{1,a} &= \sin \bm{G}_a \cdot \bm{r}_i, \\
    g_{i}^{2,a} &= \cos \bm{G}_a \cdot \bm{r}_i,
\end{align}
\end{subequations}
where $a = 1, 2$ enumerates the primitive reciprocal vectors $\bm{G}_a$ of the simulation unit cell. Additionally, we introduce the periodic distance between the particles\cite{Cassella2023-DiscoveringPhaseTransitions}, which is used to produce a periodic version of the Jastrow factor; see App.~\ref{app:NN-details} for the details. 

For the neural-network input, we combine $g_{i}^{m,a}$ (index values $m=1,2$, $a=1,2$) into a four-dimensional vector $\bm{g}_i$. It is then mapped to the embedding space (dimension $d$) by a linear transform $W_{\text{emb}}$ of dimension $d \times 4$:
\be
    \bm{h}_i^{(0)} = W_{\text{emb}} \bm{g}_i + \bm{b}. 
\ee
Embedding vectors $\bm{h}_i^{(0)}$ of each particle $i$ enter the residual stream of the Transformer. 

The backbone of the neural network architecture is a Psiformer-type\cite{Psiformer} transformer network consisting of $n_L$ layers comprising multi-head scaled dot-product attention (MHSA) and multilayer perceptron (MLP), see the details in App.~\ref{app:NN-details}. Importantly, we use a permutation-equivariant Transformer, which means that the stream of each particle $i$ is updated by the information from the set of the other particles $\{ \mathbf{r}_{/i} \}$ in a way that does not change under any permutation of $\{ \mathbf{r}_{/i} \}$. That will be important to ensure bosonic statistics of the wavefunction. 
To condition the model output on the interaction strength $\lambda$, we introduce into each Transformer layer coupling to the learned interaction token $\bm{p}_{\lambda}$. The latter is obtained once per network using a small MLP:
\be
    \bm{p}_\lambda = W_2 \tanh \left( W_1 \lambda + b_1 \right) + b_2.
\ee

The output of Transformer is used to generate $D$ generalized orbitals $\phi_d$, which are  permutation-equivariant. It means that each orbital is a function of $N$ coordinate vectors, which depends on the first argument arbitrarily while any permutation $\sigma$ of the remaining $N - 1$ arguments does not change its value. The network produces the value of each of $D$ orbitals with $i$th particle coordinate as the first argument, $\phi_d (\mathbf{r}_i, \{ \mathbf{r}_{/i} \})$. Permutation-invariance reads:
\be
    \phi_d (\mathbf{r}_i, \{ \mathbf{r}_{/i} \})  = \phi_d (\mathbf{r}_i, \sigma \{ \mathbf{r}_{/i} \}).
\ee
The values of the $\lambda$-conditioned orbitals are obtained through $D$ independent linear transforms, $W_{\text{orb}}^d$, applied to the final residual stream value $\bm{h}_i^{(n_L)}$ for each particle:
\be
    \log \phi^d (\bm{r}_i, \{\bm{r}_{/ i} \}, \lambda) = W_{\text{orb}}^d \bm{h}_i^{(n_L)}.
\ee

Finally the many-body wavefunction of the system is obtained as follows:
\be \label{output-wavefunction}
    \psi_{\lambda} (R) = J_{\lambda}(R) \sum_{d=1}^D \prod_{i=1}^N  \phi^d(\bm{r}_i, \{\bm{r}_{/ i} \}, \lambda) ,
\ee 
where $J(R)$ is the inter-particle Jastrow prefactor \eqref{Jastrow}. The logarithms of the orbital values are summed within a head and then the standard log-sum-exp trick is used to combine the heads.
Due to the equivariance of $\phi_i^d$, the output \eqref{output-wavefunction} obeys bosonic statistics being invariant under arbitrary permutations of the particle coordinates. Importantly, if the orbitals $\phi_i^d$ would depend only on $\bm{r}_i$ and not on the coordinate of the other particles (which would happen in case of the attention turned off), then their product would correspond to an ideal Bose condensate in the state $\phi^d$. Using the full generalized orbitals $\phi^d(\bm{r}_i, \{\bm{r}_{/ i} \}, \lambda)$ allows to go beyond that, but retains the capacity to effectively represent the superfluid phase. 

That concludes the description of the neural network architecture used in the present work. Its practical JAX implementation can be found in the repository~\cite{bose-repo}, which was built on top of the Psiformer code~\cite{ferminet_github, Psiformer}. The parameters of the network are given in Tbl.~\ref{tbl:parameters} of App.~\ref{app:NN-details}. 

\begin{figure*}
    \centering
    \includegraphics[width=0.99\linewidth]{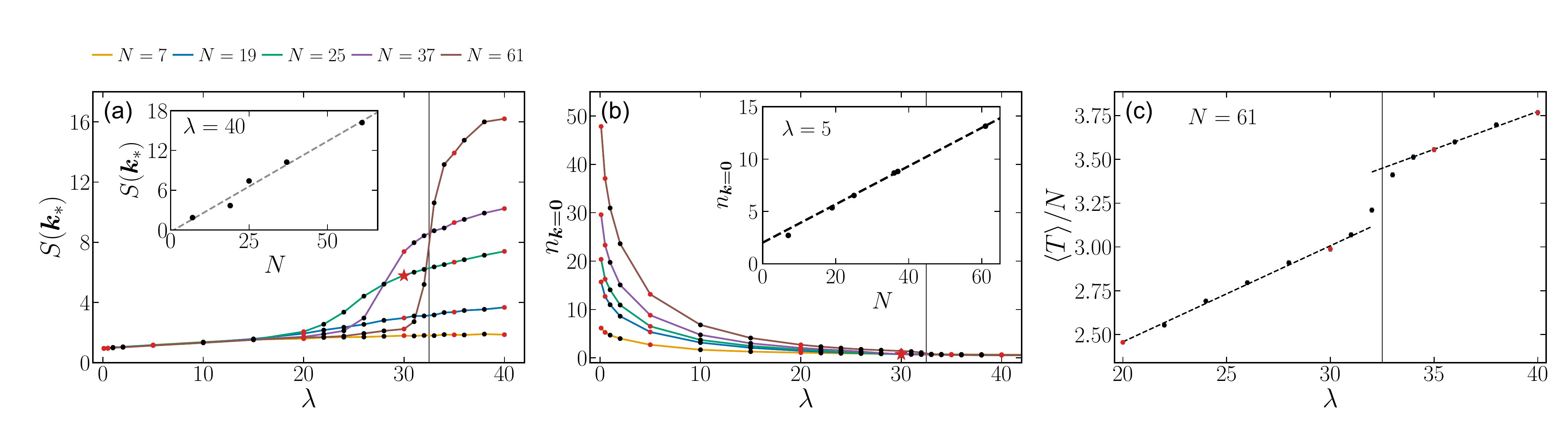}
    \caption{(a) Bragg peak height, $S(\bm{k}_*)$, as a function of interaction strength for various particle numbers $N$. The red star represents the result of a separate, single-$\lambda$ optimization demonstrating agreement with the parametric model. Inset: scaling of the Bragg peak height with $N$ at $\lambda = 40$, showing the linear dependence $S(\bm{k}_*) \propto N$ as expected in the crystal phase. 
(b) Condensate occupation, $n_{\bm{k} = 0}$, as a function of interaction strength, $\lambda$, across the phase transition for various number of particles $N$. Inset: linear scaling of $n_{\bm{k} = 0}$ with number of particles at $\lambda = 5$, indicating superfluid phase.
(c) Kinetic energy of the system as a function of interaction strength $\lambda$ at $N = 61$. The dependence undergoes a jump near the superfluid-to-crystal transition. Dashed lines are linear fits inside each phase away from transition.  
Across the panels, red dots indicate the values of $\lambda$ used during the network training ($\Lambda_{\text{tr}}$ and $\Lambda_{\text{ft}}$), while the black dots are model predictions with no re-optimizing. Thin vertical lines at $\lambda = 32.5$ indicate the approximate phase transition position at $N=61$ according to \textit{Boseformer}, which is consistent across the three panels.
    }
    \label{fig:lambda-panels}
\end{figure*}

\section{Training procedure}

\begin{figure}
    \centering
    \includegraphics[width=0.95\linewidth]{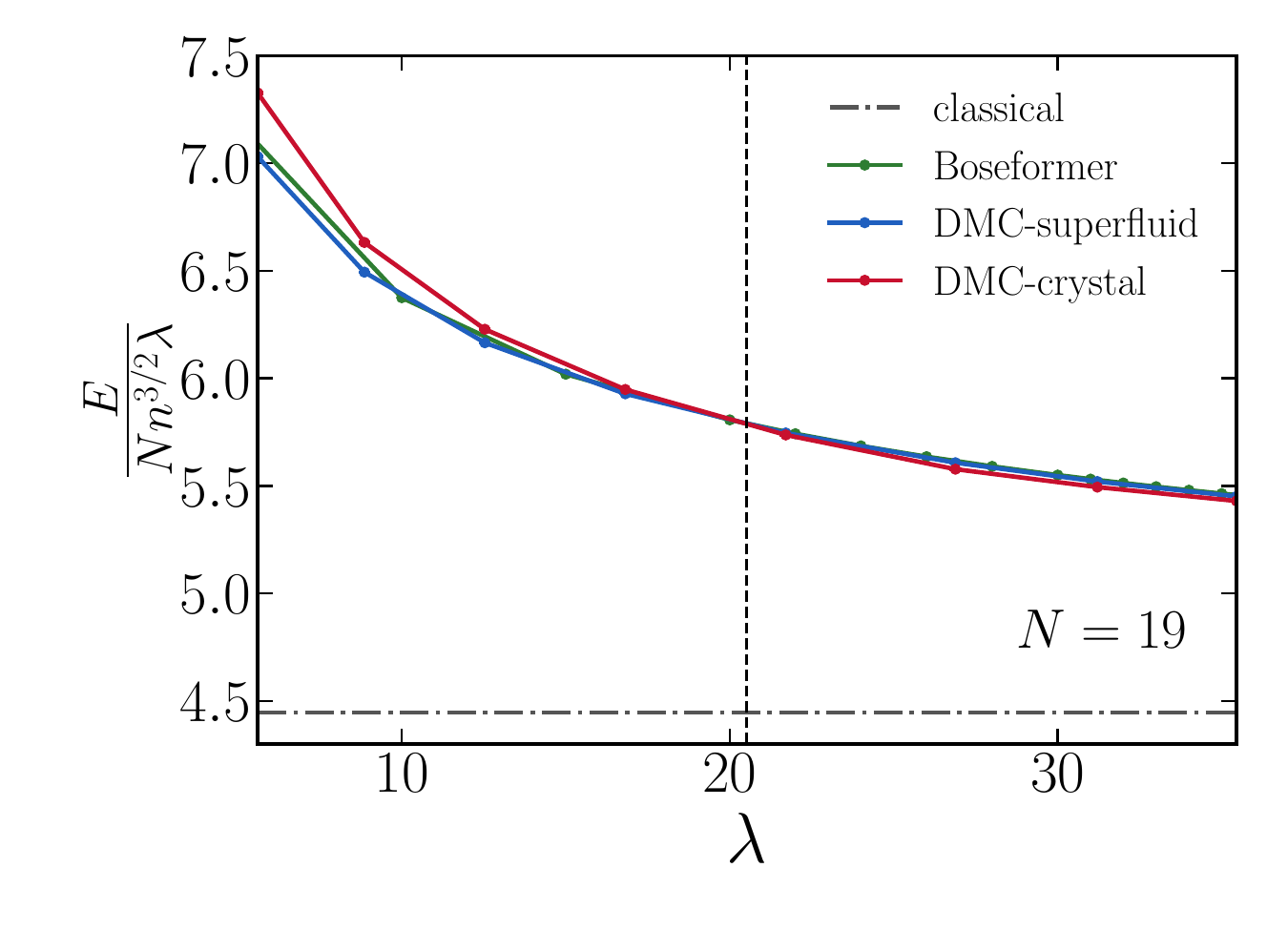}
    \caption{Ground state energy of the system with $N=19$ bosons as a function of interaction strength, $\lambda$. Variational Monte-Carlo estimate of \textit{Boseformer} is shown in green. It is compared to the predictions of the Diffusion Monte Carlo (DMC) method, reproduced after Ref.~\onlinecite{Astrakharchik}; blue (red) lines show DMC energies in the superfluid (crystal) phases, correspondingly. The methods agree within $1\%$ accuracy. Black dot-dashed line is the classical energy of the crystal while the vertical dashed line is the intersection of superfluid and crystal DMC branches. 
    }
    \label{fig:energy}
\end{figure}

In this section, we describe the unsupervised training procedure of the  neural network introduced above. 
We train one $\lambda$-parametric model for each number of particles $N=7,19,25,37$, and $61$. 
For optimizing the network parameters, we standardly use variational Monte Carlo method. Since we were aiming at training a parametric model taking interaction strength $\lambda$ as one of the inputs, the training is done simultaneously for a set of interaction strengths $\Lambda_{\text{tr}}$. The loss function reads:
\be
    \mathcal{L} = \frac{1}{|\Lambda_{\rm tr}|} \sum_{\lambda\in\Lambda_{\text{tr}}} E_\lambda^{\text{var}} \rightarrow \min_{\{\theta\}},
\ee
where $\{\theta\}$ are all network parameters and the variational energy for each $\lambda$ is given by
\be
    E_\lambda^{\text{var}} = \frac{\langle \psi_{\lambda}(\theta) | \mathcal{H}(\lambda) | \psi_{\lambda} (\theta) \rangle}{\langle \psi_{\lambda}(\theta) | \psi_{\lambda} (\theta) \rangle} .
\ee
If the neural network was capable of representing the ground state at each $\lambda$ exactly, the loss minimum would correspond to an ideal representation of the ground state for all $\lambda$ in $\Lambda_{\text{tr}}$. In our variational approach, we expect the network to learn approximate representations of wavefunctions for a range of $\lambda$, which may include phase transitions. 

We choose a set of $\lambda$ (see Tbl.~\ref{tbl:parameters}) so that it includes points on both sides of the phase transition. The resulting network is already capable of producing both superfluid and crystal wavefunction. However, for increased accuracy, we find it useful to fine-tune the resulting network across a narrower region in the vicinity of the phase transition, $\Lambda_{\text{ft}}$. For  inference, we chose either base or fine-tuned model depending on whichever has the lowest variational energy. Additionally, we trained a single-$\lambda$ model to check the performance of the conditioned model compared to the setup, in which optimization is done for each $\lambda$ separately. 

General details of the variational Monte Carlo optimization follow Ref.~\onlinecite{Psiformer}.
Monte-Carlo walkers are sampled using Metropolis Markov chain Monte Carlo (MCMC)~\cite{Metropolis1953} from the density provided by  $|\psi_{\lambda}(R)|^2$, where $\psi_{\lambda}$ is taken from the neural-network output. Since each $\lambda$ corresponds to its own wavefunction, we divide all walkers into groups per each value of $\lambda$ in the training set. For outlier control, we used standard clipping and re-centering protocol, but applied it per group with a fixed $\lambda$ since the energies of the configurations with different $\lambda$ may differ significantly. 

Second-order KFAC optimizer~\cite{Martens2015-KFAC} is used for the optimization of the model parameters $\{ \theta \}$. A decreasing learning rate $\eta(t) = 0.02 (1 + t/t_0)^{-\gamma}$ with $\gamma=1.5$ was implemented. Further details on the parameter values can be found in Tbl.~\ref{tbl:parameters}. The optimization showed that the loss function consistently decreases during the optimization and then levels off, see Fig.~\ref{fig:loss} in the Appendix. We study the obtained model in the following section and show that it successfully predicts superfluid phase at small $\lambda$ and crystal at large $\lambda$ \footnote{While for most $N$ we observed that the model converged to the commensurate crystalline order at large $\lambda$, we also observed that for $N=36$ the system sometimes converges to the $N=37$ state with one vacancy. Subsequent fine-tuning to a narrower parameter region fixes this issue.}.


\section{Observables across the phase transition}
\label{sec:observables}

In this section, we describe the physical quantities computed from the trained network to identify and characterize the quantum order it predicts at each interaction strength.
Since our method explicitly produces the ground-state wavefunction, computing observables is straightforward and amounts to a Monte Carlo inference run. Monte-Carlo procedure is the same as during training, except that the network parameters are not updated. For a fixed value of the interaction $\lambda$ supplied as network input, we run only the MCMC sampling for a few thousand steps, drawing configurations from $|\psi_{\lambda}(\bm{r})|^2$ and evaluating the observables on them. In this way, we obtain evidence of the superfluid--crystal phase transition in multiple observables, all computed from a single network spanning both phases.

\subsection{Energy and its partition into kinetic and potential parts}

\begin{figure}
    \centering
    \includegraphics[width=0.95\linewidth]{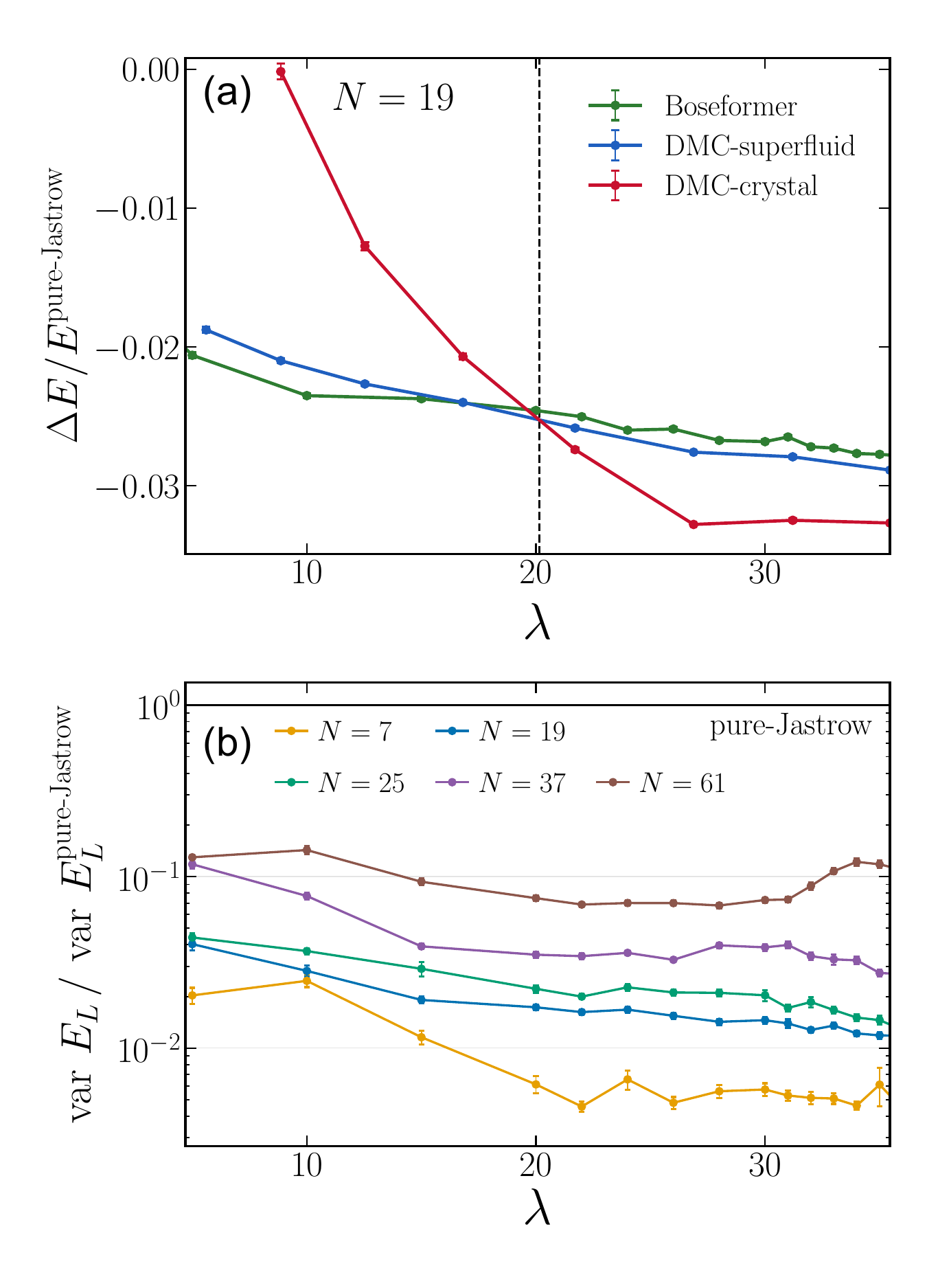}
    \caption{(a) Zoom-in of the variational energy difference $\Delta E$ between our method and one-parameter pure-Jastrow ansatz \eqref{Jastrow}, shown in green. \textit{Boseformer} is compared to the energies of the superfluid (blue) and crystal (red) branches obtained via Diffusion Monte Carlo (DMC) method.
    (b) Local energy variance (see Sec.~\ref{sec:v-score}), normalized by the one of the pure-Jastrow wavefunction baseline. 
    }
    \label{fig:energy-and-variance}
\end{figure}

The variational Monte-Carlo energy was used as the loss function during network training. In Appendix~\ref{app:NN-details}  (Fig.~\ref{fig:loss}), we show its dependence on the training iteration, which demonstrates monotonic decrease and saturation. 

In Fig.~\ref{fig:energy} (green curve), we plot the resulting variational energy as a function of interaction strength, $\lambda$, for a system with $N = 19$ bosons. It is obtained from the trained model by running inference MCMC chain (no network parameter updates) and averaging over the last few thousand iterations after equilibration. In the figure, we normalize the energy by $N n^{3/2} \lambda$ motivated by the expression for the classical energy of the crystal phase, $E_{\text{cl}} = 4.45 N n^{3/2} \lambda$, which is shown by a dash-dotted line. We observe that the scaled energy $E / (N n^{3/2}\lambda)$ decreases with $\lambda$ remaining above the classical crystal energy. 

We benchmark the energy obtained by our method against the Diffusion Monte Carlo (DMC) method.  The results of our DMC calculation, reproducing Ref.~\onlinecite{Astrakharchik}, are shown for the system  in Fig.~\ref{fig:energy} in blue (superfluid branch) and red (crystal branch). For $N = 19$ particles,  super-fluid-to-crystal first-order phase transition happens at $\lambda = 20.5$, which differs from the $N\to\infty$ extrapolation value of $\lambda = 30$. 
Contrary to DMC, our method does not produce two superfluid and crystal branches separately but is  trained to directly generate the most energetically favorable phase at each $\lambda$. The results demonstrate that in general the energies obtained by our method match DMC calculation with the accuracy of around $1\%$.
In the Fig.~\ref{fig:energy-and-variance}(a) we zoom into the transition region, normalizing by the baseline energy of the pure-Jastrow wavefunction $\psi(R) = J_{\lambda}(R)$, a minimal interaction-aware ansatz for the superfluid phase with a single optimized variational parameter $R_0$. 
We see that the agreement between DMC and \textit{Boseformer} is better in the superfluid phase, while in the crystal phase \textit{Boseformer} energy is slightly higher and is closer to the superfluid branch of DMC. Nevertheless, as the structure factor shows, \textit{Boseformer} successfully describes reconfiguration of the ordering into a crystal. While \textit{Boseformer} energy prediction is close, it is also able to directly generate the wavefunctions that can be used to compute any observables. 

The superfluid-to-crystal phase transition does not visually show up in the dependence of the total energy on $\lambda$, obtained from  \textit{Boseformer}. However, the phase transition is visible on the separate plots of the kinetic or potential parts of the total energy. In Fig.~\ref{fig:lambda-panels}(c), we show the kinetic energy per particle, $\langle T \rangle / N$ as a function of $\lambda$ for $N=61$ particles in the system. As one can see from the plot, at $\lambda \approx 32$ kinetic energy undergoes a jump, indicating a phase transition.

\subsection{Pair correlator}

Next, we study the pair correlator of bosons, $g_2(\bm{r})$. It is well suited to observe the transition to a crystal phase and should be used instead of the MCMC-averaged plain particle density $\rho(\bm{r})$. Indeed, the latter may remain uniform in the crystal phase due to translational invariance in the periodic simulation cell. Contrary to that, $g_2(\bm{\br})$ demonstrates the crystal structure even if one averages over random shifts of the emergent crystal. We define it as
\be
    g_2(\bm{\br}) = \frac{|A|}{N(N - 1)}\left\langle \sum_{i \neq j} \delta(\bm{r} - \bm{r}_{ij}^{\min}) \right\rangle ,
\ee
where $A$ is the area of the simulation cell, $\delta(\bm{r})$ is the periodic delta-function on the simulation torus, and $\bm{r}_{ij}^{\min}$ is the minimum-image distance between particles $i$ and $j$. The normalization corresponds to $g_2(\bm{r}) = 1$ for uncorrelated particles of the ideal gas. In practice, $g_2(\bm{\br})$ is computed as a histogram of the $\bm{r}_{ij}^{\min}$ in the last iteration of MCMC inference run.

The results for a number of boson numbers $N$ and interaction strengths are presented in Fig.~\ref{fig:g2-maps}(b) as a table with different values of $\lambda$ and $N$. Quite demonstrably, the plot demonstrates superfluid-to-crystal transition with the increase of $\lambda$, which gets sharper as $N$ grows. In the superfluid the only feature is a dip in $g_2(\bm{r})$ at $\bm{r} = 0$ due to boson-boson repulsion. As $\lambda$ grows, first one can see isotropic concentric density modulations, which then acquire a hexagonal warping and granularity before developing into a clearly seen triangular-lattice pattern at large $\lambda$. In the next subsection, we quantitatively describe this process by studying the structure factor.

\subsection{Structure factor and Bragg peaks}
\label{sec:structure-factor}

As the next observable, we compute the static structure factor $S(\bm{k})$, i.e. the density-density correlator in the Fourier space. It can be estimated as the following average over Monte Carlo walkers:
\be
    S(\bm{k}) = \frac{1}{N} \left\langle \left|\sum_{j=1}^{N} e^{i \bm{k} \bm{r}_j} \right|^2 \right\rangle .
\ee
The quasi-momentum $\bm{k}$ takes values among the simulation-cell reciprocal lattice vectors, $\bm{G}_{ij} = i \bm{G}_1 + j \bm{G}_2$. 
In the presence of crystalline order, the structure factor develops Bragg peaks determined by the crystal geometry and lattice constant. The crystals that emerge in our setup have triangular lattice and are commensurate with the simulation unit cell. Therefore, there are six main Bragg peaks, which are equidistant to the $\bm{k} = 0$ point and related by a $C_6$ rotation symmetry. 
To estimate the height of the main Bragg peak, $S(\bm{k_*})$, we search for the highest maxima in the $S(\bm{k})$ on the first commensurate triangular reciprocal shell and average over the six values at $\bm{k}$-points obtained by $C_6$-rotations around origin [so formally $S(\bm{k}_*)^{\text{estim.}} = (1/N_{\mathcal{R}}) \sum_{\mathcal{R}} S(\mathcal{R}\bm{k}_*)$]. 

The results for $S(\bm{k}_*)$ are presented in Fig.~\ref{fig:lambda-panels}(a) for different particle number $N$ and demonstrate a clear phase transition behavior---$S(\bm{k}_*)$ undergoes a sharp increase at $\lambda$ in the range $20$--$40$ for sufficiently large $N$. Importantly, the sharpness of the transition and $S(\bm{k}_*)$ value at large $\lambda$ grow with $N$. The latter is indicative of the crystal phase, where $S(\bm{k}_*)$ value is extensive and should grow linearly with $N$. To check the scaling, we plot the dependence of $S(\bm{k}_*)$ on $N$ for $\lambda = 40$ in the inset to Fig.~\ref{fig:lambda-panels}(a), which indeed demonstrates a linear dependence. 

\subsection{Condensate occupation $n_{k = 0}$}

Next, we estimate the zero-momentum condensate occupation $n_{\bm{k}=0}$, which represents the
off-diagonal long-range order of the superfluid. It can be expressed through the zero-momentum Fourier harmonic of the one-body density matrix~\cite{PenroseOnsager1956,Giorgini1996},
\begin{align}
    \rho_1(\bm{r}, \bm{r}') & = \frac{N}{\langle \psi | \psi \rangle}
    \int d\bm{r}_2 \dots d\bm{r}_N \,
    \\ \nonumber
    & \qquad
    \psi^*(\bm{r}, \bm{r}_2, \dots, \bm{r}_N)\, \psi(\bm{r}', \bm{r}_2, \dots, \bm{r}_N).
\end{align}
Substituting $\bm{r}' = \bm{r} + \bm{s}$ and factoring out $|\psi(R)|^2$, we obtain
\be
    n_{\bm{k}=0} = \frac{N}{A} \int d\bm{s} \left\langle \frac{\psi(R_{\bm{s}})}{\psi(R)} \right\rangle,
\ee
where $R$ is the array of particle coordinates,
$R_{\bm{s}} = (\bm{r}_1 + \bm{s}, \bm{r}_2, \dots, \bm{r}_N)$ and $\langle \dots \rangle$ denotes the average
over $R \sim |\psi(R)|^2$. We evaluate it over the Monte Carlo walkers, sampling the shift vector
$\bm{s}$ uniformly over the cell.

We show the results for $n_{\bm{k} = 0}$ in Fig.~\ref{fig:lambda-panels}(b). We observe that the condensate occupation is non-zero at small $\lambda$ and scales linearly with the number of particles $N$, which signifies the superfluid phase. At $\lambda \approx 25$--$35$ that condensate occupation plummets and remains strongly depleted at larger $\lambda$ consistent with the crystal phase. The obtained results show no evidence of a supersolid phase with the coexistence of superfluidity and crystallization.

\subsection{Variance of local energy}
\label{sec:v-score}

Here we study the fluctuations of the local energy, $E_L = \mathcal{H} \psi / \psi$, in order to characterize the convergence of the variational method. While the average of $E_L$ over the distribution defined by a variational wavefunction, $|\psi(R)|^2$, produces the average of the Hamiltonian over this state, $\langle E_L \rangle = E^{\text{var}}$, the variance of $E_L$ provides information on how good the variational wavefunction is. Indeed, for the true ground state (or any Hamiltonian eigenstate with the energy $E$), $E_L = E$ identically and has zero variance. For this reason, the quality of the variational approximation is often characterized by the V-score\cite{Wu2024-VScore, NQS-ScalingLaws}, $V_0 = N\,\text{var}(E_L) / \langle E_L\rangle^2$, which uses the normalization of the variance by the variational energy with an additional factor $N$ needed for the non-extensive thermodynamic limit. 

However, in our case the ground state energy of the superfluid vanishes at $\lambda\to0$, so normalization by the energy is not informative in this limit. Instead, we opt to normalize the variance by that of the pure-Jastrow wavefunction $\psi(R) = J_{\lambda}(R)$, which has been used above as a baseline for the variational energy. We present the result for $\text{var } E_L / \text{var }E_L^{\text{pure-Jastrow}}$ in Fig.~\ref{fig:energy-and-variance}(b). We observe our method shows the variance of the local energy 1-3 orders of magnitude smaller compared to the baseline, signifying a much better approximation.

\section{Discussion}

In the current work, we presented a general neural quantum wavefunction for interacting bosons and showcased its application to a paradigmatic system of bosons with $1/r^3$ repulsion~\cite{Mora2007}. Our results, on the one hand, demonstrate the validity of our method and neural-network architecture, while, on the other hand, provide an independent confirmation of the quantum phase transition from superfluid to crystal, predicted in Ref.~\onlinecite{Astrakharchik}. Experimentally, this physical model may be applied to the interlayer excitons with out-of-plane dipole moments in TMD materials \cite{BermanKezerashvili2016,Li2020DipolarExcitons}. 

We compared the performance of our model to the Diffusion Monte Carlo, which is a powerful method capable of converging to the ground state for systems containing up to $N\sim10^3$ particles but has its own limitations: it does not provide an explicit representation of the many-body wavefunction, and natural estimates for observables are unbiased for energy and observables commuting with the Hamiltonian only.
In order to study the superfluid-to-crystal transition with DMC, one has to use specially prepared trial wavefunctions, which bias the method toward one of the two phases. In contrast, \textit{Boseformer} is trained to predict the optimal wavefunction directly at each $\lambda$. It successfully captures the phase transition and provides a universal ansatz that works across it. At the first-order transition, the jump of the wavefunction is smeared into a steep crossover, obtained from a single pretrained network. The crossover becomes sharper as the number of particles increases. While the network capacity likely imposes some limit on the sharpness of the transition even in the thermodynamic limit, the current behavior is sufficient to observe and study the phase transition.

One of the intriguing theoretical possibilities in bosonic systems is the existence of the supersolid phase, combining non-zero superfluid stiffness and crystalline order. Previous research \cite{Kurbakov2010} concluded that no such phase emerges for the Hamiltonian \eqref{Hamiltonian} in the absence of additional perturbations. Our  study did not reveal any supersolid state confirming that statement. However, extension to the systems with larger $N$ is required to reliably refute the supersolid hypothesis within our approach. More generally, further directions of research may include scaling the computational simulation to systems with larger number of particles and applying the method to various bosonic Hamiltonians.

\section*{AI disclosure}

The initial implementation of the \textit{Boseformer} architecture was done by one of the authors (D. A.) with no AI coding assistance. In the following, various Claude and ChatGPT models were used for further development, running, and analyzing the results of \textit{Boseformer}. The authors take responsibility for the scientific content and the writing. 

\section*{acknowledgements}

D. A. is grateful to Khachatur Nazaryan and Timothy Zaklama for valuable discussions and comments. This work was supported at Yale University by NSF Grant No. DMR-241018.


\appendix

\section{Details of the neural-network architecture and optimization procedure}
\label{app:NN-details}

Here we provide details on the architecture of the neural network that we used in our work. The specific parameter values used in our work can be found in Tbl.~\ref{tbl:parameters}.

\subsection{Periodic interparticle distance}

The Jastrow prefactor \eqref{Jastrow} in the variational wavefunction that we use in our work is defined through the periodic distance, 
\be
    \tilde{r}_{ij} = \frac{1}{2\pi} \sqrt{\bm{\alpha}^T \Gamma \bm{\alpha} + \bm{\beta}^T \Gamma \bm{\beta}},
\ee
where the metric $\Gamma = A^T A$ is expressed through the real-space lattice vectors of the simulation unit cell, $A = [\bm{a}_1 | \bm{a}_2]$, while the phase components are defined through the reciprocal vectors $\bm{G}_a$:
\begin{align}
    \alpha_a &= 1 - \cos \bm{G}_a \cdot (\bm{r}_i - \bm{r}_j), \\
    \beta_a &= \sin \bm{G}_a \cdot (\bm{r}_i - \bm{r}_j).
\end{align}
The periodic distance $\tilde{r}_{ij}$ approximates the interparticle distance $|\bm{r}_i - \bm{r}_j|$ when it is small compared to the size of the periodic unit cell, while being smooth and periodic in this unit cell for arbitrary values of the arguments.
We also introduce a capped periodic distance
\be
    \rho_{ij} = R_0 \tanh \frac{\tilde{r}_{ij}}{R_0},
\ee
where $R_0 = \zeta  L$ is the parameter that regulates the saturation of $\rho_{ij}$ away from the cusp region. The `matching fraction' $\zeta$ is fixed in \textit{Boseformer}, see Tbl.~\ref{tbl:parameters}. As mentioned above, the values of $\rho_{ij}$ are used in the Jastrow prefactors, Eq.~\eqref{Jastrow} but do not enter the neural network. 

\subsection{Transformer architecture details}

The central block of our neural-network architecture is an  equivariant Psiformer-type\cite{Psiformer} transformer network. It has $H$ heads, which satisfy the relation $d = H d_k$, where $d$ is the residual stream dimension and $d_k$ is the dimension of each head.
Each particle retains its own residual stream $\bm{h}_i$ and gathers information about the other particles through attention only. The unordered set of all residual-stream values excluding the stream of a selected particle is denoted as $\{ \bm{h}_{/i} \}$. The weights of the attention and MLP blocks are shared across particles but each layer has
its own attention and MLP weights. Together, these properties ensure equivariance: for every particle, any permutation of the remaining particles does not change its residual stream.  Thus, attention represents communication between the particles while MLP does computation inside each stream. The flow is as follows:
\begin{align}
     \bm{h}_i & \rightarrow \text{LN}\left( \bm{h}_i + \text{MHSA}(\bm{h}_i,\{ \bm{h}_{/i} \}) \right), \\
    \bm{h}_i & \rightarrow \text{LN} \left( \bm{h}_i + \text{MLP}(\bm{h}_i) \right), \\
    \bm{h}_i & \rightarrow  \bm{h}_i + \tanh W_l \bm{p}_\lambda, \label{coupling-to-lambda-token}
\end{align}
where LN is post-residual LayerNorm~\cite{Ba2016-LayerNormalization} that normalizes each particle's residual stream. 
MHSA computes per-head attention outputs
\be
    a_i^k = \sum_{j} \text{softmax}_j \left[\frac{(W_Q^k \bm{h}_i) \cdot (W_K^k \bm{h}_j)}{\sqrt{d_k}}\right] W_V^k \bm{h}_j,
\ee
using the query, key and value matrices ($W_{Q}$, $W_{K}$, and $W_{V}$) of dimensions $H d_k \times d$, comprising blocks corresponding to each head ($W_{Q}^k$, $W_{K}^k$, and $W_{V}^k$). The resulting head outputs are concatenated and projected back to the residual stream dimension with the matrix $W_O$
\be
    \text{MHSA}(\bm{h}_i,\{ \bm{h}_{/i} \}) = W_O [a_i^1, \dots, a_i^H],
\ee
and then added to the stream. 
MLP blocks are represented by a two-layer \texttt{tanh}-activation perceptron restricted to the individual stream of each particle.
An important elaboration on top of Psiformer is the inclusion of the operation \eqref{coupling-to-lambda-token}, which introduces model conditioning on the Hamiltonian parameter: interaction strength $\lambda$.  As explained in the main text, interaction token $\bm{p}_{\lambda}$ is obtained by a trainable linear transform and shared across the network. On the practical side, JAX's fused function \texttt{jax.nn.dot\_product\_attention} is used for efficiency in the attention part. 

\begin{figure}
    \centering
    \includegraphics[width=0.95\linewidth]{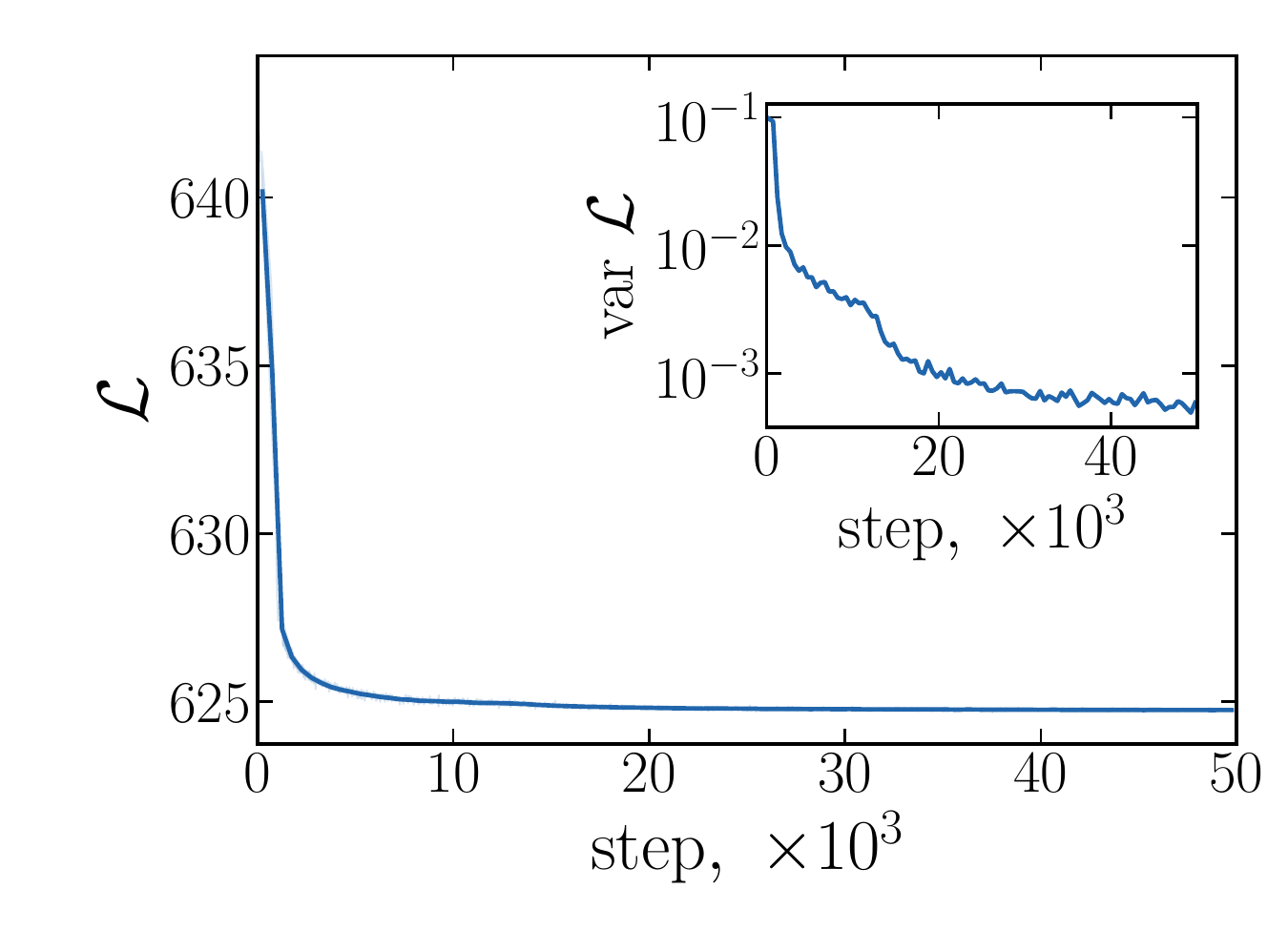}
    \caption{Loss function during training for $N=25$ particles (mean local energy over walkers). It demonstrates monotonic decrease and saturation. Inset: running variance of the batch-mean loss across iterations, with weights that decay exponentially with the age of each iteration. It drops by two orders of magnitude during the optimization.}
    \label{fig:loss}
\end{figure}

The parameters of the neural network and the optimizer are presented in Tbl.~\ref{tbl:parameters}. The typical behavior of the loss function during optimization demonstrates a monotonic convergence; see Fig.~\ref{fig:loss}.

\begin{table}
  \caption{
  Parameters of the \textit{Boseformer} neural network and the optimizer used in our work.}
      \label{tbl:parameters}
      \begin{ruledtabular}
      \begin{tabular}{lll}
      Group & Parameter & Value \\
      \hline \hline
        Backbone & embedding dim. $d$ & 256 \\
       & N of layers $n_L$ & 4 \\
       & N of att. heads $H$ & 4 \\
       & head dim. $d_k$ & 64 \\
       & MLP hidden dims. & (256,) \\
       & N of orbitals $D$ & 16 \\
      \hline
      $\lambda$-cond. & $\lambda$-embed. hidden dim & 64 \\
       & dim. of $\lambda$ token $(\bm{p}_\lambda$) &  256 \\
      \hline
      Jastrow & matching fraction $\zeta$ & 1.0 \\
      \hline
      & Total network params.  & $1.9 \cdot 10^6$  \\
      \hline
      \hline
      Ewald sum. & N of real shells & 4 \\
       & N of reciproc. shells & 12 \\
      \hline
      Multi-$\lambda$  & training set $\Lambda_{\text{tr}}$ & [0.1, 0.5, 5, 20, 50, 80] \\
       & fine-tune set $\Lambda_{\text{ft}}$ & [20, 25, 30, 35, 40] \\
      \hline
      MCMC & sampler & RW-Metropolis \\
       & Metropolis steps/iter. & 30 \\
       & initial step width  & 0.02
  \\
        & adapt. frequency & 100
  \\
       & burn-in steps & 300 \\
      \hline
      Optimiser & algorithm & KFAC \\
       & N of iterations & $5\cdot10^4$ (base),  $2\cdot10^4$ (ft) \\
       & Batch size & 2048--4096 \\
       & learning rate & 0.02 \\
       & learning rate delay $t_0$ & $10^4$ (base), $3\cdot10^4$ (ft) \\
       & lr decay exponent $\gamma$ & 1.5 \\
       & Local energy clipping & 3.0 \\
       & KFAC damping & 0.01 \\
       & KFAC norm constraint & 0.003 \\
      \end{tabular}
      \end{ruledtabular}
\end{table}

\section{Benchmarking against Diffusion Monte Carlo (DMC)}
\label{sec:DMC}

As shown in Sec.~\ref{sec:observables}, we compare our method against Diffusion Monte Carlo (DMC), which is sign-free for bosons in the absence of a magnetic field. DMC is a standard choice for this system, and its main advantage is that it converges to the exact ground-state energy in the limit of an infinite number of Monte Carlo walkers and an infinitesimal time step $\Delta \tau$. However, DMC does not provide the ground-state wavefunction explicitly. In general, its standard mixed estimators are biased for observables other than energy or operators commuting with the Hamiltonian.

We follow Ref.~\onlinecite{Astrakharchik} and simulate $N_W = 2048$ Monte Carlo walkers in imaginary time using importance sampling\cite{Reynolds1982}. Likewise, for each $\lambda$, we use separate guiding wavefunctions and initial configurations to track the liquid and crystal branches. For the superfluid, we use a Jastrow guiding wavefunction based on \eqref{Jastrow}, while for the crystal, there is an additional Gaussian factor that pins the particles to the triangular lattice sites. To match the \textit{Boseformer} setup, we use the same rhombic simulation cell, density, particle numbers $N$, and Ewald summation procedure\cite{Mora2007}. We use a time step $\Delta \tau \sim 10^{-5} \lambda^2$; we checked that finite $\Delta \tau$ introduces an energy bias that is linear in $\Delta \tau$. One can use $\Delta \tau \to 0$ extrapolation to improve the accuracy of the DMC method.

Importantly, the intersection of the superfluid and crystal branches depends strongly on the particle number $N$. For the \textit{Boseformer}-relevant $N=19$--$61$, the fitted branch intersections are located at $\lambda = 18$--$21$, while the $N\to\infty$ extrapolation (which is done by fitting to the $E_N / N = E_{\infty} + C/\sqrt{N}$ dependence with $N$ up to $\sim10^3$) reproduces $\lambda_c \approx30$, matching Ref.~\onlinecite{Astrakharchik}.

As we show in the main text (see Fig.~\ref{fig:energy}), our method converges to variational energies typically within $1\%$ of DMC for particle numbers $N = 7$--$61$. At the same time, it provides the full many-body wavefunction over a range of Hamiltonian parameter values from a single model.

\newpage

\bibliography{NQS-bosons}

\end{document}